# Health Information Search Behavior on the Web: A Pilot Study

Shanu Sushmita, PhD, Si-Chi-Chin, PhD,
University of Washington-Tacoma, WA

**Abstract**

*Searching health information on web has become an integral part of today's world, and many people turn to the Web for healthcare information and healthcare assessment. Our pilot study investigates users' preferences for the **type** of search results (image, news, video, etc.), and investigates users' ability to accurately interpret online health information for the purpose of self-diagnosis. The preliminary results reveal that blog and news articles are most sought by users when searching online information and there exist challenges in the use of online health information for self-diagnosis.*

**Introductions**

With the growth of search engines, accessing health information on the World Wide Web became easier and faster. A recent study reported that 80% of web users have browsed online for health information. Due to this increased usage of health information among people, there is a growing interest among researchers and search engine companies to understand users' health information seeking behavior. Although several efforts[3-7] have been made, there remain many unanswered questions in this area of study. The goal of our study is to explore the following research questions: **(R1)** *What types of search results (image, news, blog, video, etc.) are relevant to health related topics?* **(R2)** *Are web users able to usefully interpret the health information on the web?*

**Experimental Design**

A within subject[8] experimental design was used in our pilot study, where we investigate: (1) participants' preference on various types of search results (image, news, blog, video, etc.); (2) participants' ability to usefully interpret online health information and identify a target disease. A total of 9 participants (mean age 26 years) performed the study, of which 4 were female and 5 male. Participants of our pilot study were asked to perform four search tasks (2 tasks for **R1,** and 2 tasks for **R2**) using their preferred search engine. Each search task was based on the simulated work task situation framework proposed by Borlund[2]. Participants were recruited through an email distributed to several mailing lists. The experiment was divided into three sessions, (i) an entry questionnaire was used to capture participants' profile and search background, (ii) post-session questionnaires were used to capture subjective assessments on the tasks (iii) Exit-questionnaires provided participants' perceptions of tasks as a whole.

**Result and Conclusion**

To answer our first research question (**R1**), participants were asked to provide their preference of result type when searching for health information, on the scale of 1-5 (1-Very little, 5- Very Much). The results show that blogs and news results (median 4) were preferred more than image (median 3) and video results (median 2). Furthermore, the study also compared the "type" of information collected by participants in response to their search tasks. The information collected by participants also echoed with the preference scores provided by them. Participants collected most information from news and blog articles. These preliminary results suggest the importance of blog and news information in online health informatics. However, the reason for such preferences is still not clear and we aim to investigate this in our future research.

The aim of the second research question (**R2**) was to investigate if users are able to use online information to identify diseases. As part of the search task (2 tasks in total), participants were given a list of symptoms to search and identify the possible disease. The results show that most users had challenges to identify the disease accurately despite the large amount of health information resource on the Web. For both tasks, only one participant was able to identify the disease based on the given symptoms. The low success rate may due factors like: searching difficulties, disorganization, information overload, etc. Lack of education in using medical terminology could be also limit users from accurately interpreting online information[10]. As a result, users may misjudge information, become information-overloaded and thereby easily get confused[9]. Through our research, we aim to investigate these factors and finding ways to present usable health information.